\documentclass[leqno,draft,11pt]{article}
\usepackage{amssymb,amsmath,theorem,a4wide}

\def\enumup{}

\ifx\mydefs\undefined\else \fi
\let\mydefs\relax

\allowdisplaybreaks[2]

\ifx\loadcyr\undefined\else
\input cyracc.def
\makeatletter
 at 1\@ptsize pt
 at 1\@ptsize pt
 at 1\@ptsize pt
\makeatother

\fi

\let\M\mathit
\def\gobble#1{}
\def\fixsup#1#2{{#1\let\dp\gobble\mathstrut}^#2_}

\def\bme{\hskip.75em\relax}

\def\nequiv{\not\equiv}

\let\TO\Rightarrow
\def\?{\mathbin?}

\newbox\circlebox
\setbox\circlebox\hbox{$\bigcirc$}
\def\circled#1{%
  \setbox0\hbox to\wd\circlebox{\hss$#1$\hss}\wd0=0pt
  \box0\copy\circlebox}

\let\fii\varphi
\let\tet\vartheta
\let\ep\varepsilon

\def\greek#1{$\expandafter\greeknum\csname c@#1\endcsname$}
\makeatletter
\def\greeknum#1{\ifcase#1\or\alpha\or\beta\or\gamma\or\delta\or\ep
      \or\digamma\or\zeta\or\eta\or\tet\or\iota\else\@ctrerr\fi}
\makeatother

\def\p#1{\langle#1\rangle}

\def\lh#1{\lvert#1\rvert}
\let\abs\lh

\let\bez\smallsetminus

\let\sset\subseteq

\let\onto\twoheadrightarrow

\def\two{\mathbf2}

\def\twoprimes{\raise.2\fontdimen6\scriptfont2\hbox{$\scriptstyle\prime\prime$}}
\newcommand\rpair[3][3em]{\mathrel{%
   \begin{matrix}%
     \strut\smash{\xrightonto{\hbox to#1{\hss$#2$\hss}}}\\[-1.7ex]%
     \strut\smash{\xleftembed[\hbox to#1{\hss$#3$\hss}]{}}%
   \end{matrix}}}
\makeatletter
\newcommand\xrightonto[2][]{\ext@arrow 0359\rightontofill{#1}{#2}}
\newcommand\xleftembed[2][]{\ext@arrow 3095\leftembedfill{#1}{#2}}
\def\leftembedfill{\arrowfill@\leftarrow\relbar\hookleftnoarrow}
\def\rightontofill{\arrowfill@\relbar\relbar\onto}
\def\hookleftnoarrow{\DOTSB\relbar\joinrel\rhook}
\makeatother

\def\ljk{\genfrac(){}{}}
\def\mljk{\genfrac[]{}{}}
\def\tljk#1#2{(#1|#2)}
\def\tmljk#1#2{[#1|#2]}

\def\fl#1{\lfloor#1\rfloor}

\def\cl#1{\lceil#1\rceil}

\mathchardef\#="2023 % \mathbin\#

\let\ass\leftarrow

\ifx\busspf\undefined\else
\usepackage{bussproofs}
\EnableBpAbbreviations

\fi

\ifx\symlasy\undefined

  \def\centerdot#1{%
    \setbox0\hbox{$\mathop{#1}$}\dimen0 \ht0
    \setbox0\hbox{$#1$}\advance\dimen0 -\ht0
    \setbox2\hbox to\wd0{\hss$\mathop{\cdot}$\hss}\wd2=0pt
    \lower\dimen0\box2\box0 }
\else
  \def\centerdot#1{%
     \setbox0\hbox{$#1$}%
     \raise0.206\ht0\hbox to\wd0{\hss$\cdot$\hss}%
     \kern-\wd0 \box0 }
\fi

\let\sls|

\def\Up{{\setbox0\hbox{$\uparrow$}%
         \lower\dp0\hbox to\wd0{\hss\vrule width4pt height.4pt\hss}%
         \kern-\wd0\box0}}
\def\UP{{\setbox0\hbox{$\uparrow$}%
         \lower\dp0\hbox to\wd0{\hss\vrule width4pt height.4pt\hss}%
         \kern-\wd0\copy0\kern-\wd0\raise.35ex\box0}}
\def\Down{{\setbox0\hbox{$\downarrow$}%
         \raise\ht0\hbox to\wd0{\hss\vrule width4pt depth.4pt\hss}%
         \kern-\wd0\box0}}

\let\Pr\relax
\DeclareMathOperator\Pr{Pr}%get rid of \displaylimits

\let\Re\relax

\DeclareMathOperator\Re{Re}

\let\cxt\mathrm
\def\np{\cxt{NP}}

\def\rp{\cxt{RP}}

\def\zpp{\cxt{ZPP}}

\def\fp{\cxt{FP}}

\def\tfrp{\cxt{TFRP}}

\def\tfzpp{\cxt{TFZPP}}
\def\ppa{\cxt{PPA}}
\def\ppad{\cxt{PPAD}}
\def\ppp{\cxt{PPP}}
\def\pwpp{\cxt{PWPP}}

\def\task#1{{\normalfont\textsc{#1}}}
\def\pig{\task{Pigeon}}
\def\wpig{\task{WeakPigeon}}
\def\lone{\task{Lonely}}
\def\fac{\task{Factoring}}
\def\ffac{\task{FullFac}}
\def\fsr{\task{FacRoot}}
\def\wfsr{\task{WeakFacRoot}}
\def\fsrm{\task{FacRootMul}}
\def\fsro{\task{FacRootOdd}}
\def\fsre{\task{FacRootEven}}
\def\sroot{\task{Root}}
\def\qrec{\task{QuadRec}}

\def\sig{\Sigma^b_}

\def\delz{\Delta_0}%^0

\def\st{\expandafter\hat}

\def\idz{I\Delta_0}

\def\pv{PV}

\def\sss{S^1_2}

\def\wphp{\M{WPHP}}

\def\coun{\M{Count}}

\def\grh{\M{GRH}}
\def\grhq{\M{GRH_q}}

\def\Z{\mathbb Z}
\def\C{\mathbb C}

\let\stm\N

\mathcode`\*="0203

\mathchardef\mhyphen="2D

\def\noproof{\leavevmode\unskip\bme\vadjust{}\nobreak\hfill$\qed$\par}
\let\qed\Box
\newenvironment{Pf}[1][]
  {\par\noindent\textit{Proof\optpar{#1}:}\bme\ignorespaces}
  {\noproof\pagebreak[2]\vskip\medskipamount\ignorespacesafterend}
\def\qedhere{\relax\ifmmode\eqno\qed\expandafter\aftergroup
                   \else\noproof\fi\noqed}
\def\noqed{\let\noproof\relax}

\theoremstyle{plain}
\ifx\shortthm\undefined
\newtheorem{Thm}{Theorem}[section]
\else
\newtheorem{Thm}{Theorem}
\fi

\newtheorem{Cor}[Thm]{Corollary}
\newtheorem{Lem}[Thm]{Lemma}

\newtheorem{Prob}[Thm]{Problem}

\def\theCl{\arabic{Cl}}

\theorembodyfont\upshape
\newtheorem{Def}[Thm]{Definition}

\newtheorem{Exm}[Thm]{Example}

\newenvironment{Pf*}{\let\qed\qedCl\Pf}\endPf

\usepackage[reftex]{theoremref}
\newif\iflinenumbers
\linenumberstrue
\newenvironment{algo}[1][20em]{\catcode`\^^I=13 \obeylines\doalgo{#1}}{}
{\catcode`\^^I=13 \catcode`\^^M=13
\gdef\doalgo#1#2\end#{\hbox to\hsize{\hss \let^^I\qquad%
  \def\\^^M{\nobreak\hfil\break\vadjust{}\qquad}%
  \fboxsep1em \linenum0 %
  \fbox{\hsize#1\vbox{%
  \everypar{\advance\linenum1 %
      \hbox to1.2em{%
           \hss\iflinenumbers$\scriptstyle\the\linenum$\hskip.6em\fi}}%
  #2}}\hss}\end}}
\newcount\linenum

\def\key{\relax\ifmmode\expandafter\mathbf\else\expandafter\textbf\fi}
\def\allowhyphens{\nobreak\hskip0pt\relax}
\def\hyph{\allowhyphens-\hskip0pt\relax}
\DeclareRobustCommand*\magiclparen{\ifmmode(\else\textup(\allowhyphens\fi}
\DeclareRobustCommand*\magicrparen{\ifmmode)\else\textup)\fi}
\let\lparen=(  \let\rparen=)
\def\magicparon{\catcode`\(\active\catcode`\)\active}
\def\magicparoff{\catcode`\(12 \catcode`\)12 }
\AtBeginDocument{\ifx\ifPreview\iftrue\else\magicparon\fi}
\magicparon
\let (=\magiclparen  \let )=\magicrparen

\ifx\enumup\undefined
  
\else
  
\fi

\def\optpar#1{\ifx\relax#1\relax\else\/ (#1)\fi}
\magicparoff

\mathchardef\comma=\mathcode`\,
{\catcode`\,=\active \gdef,{\comma\penalty\relpenalty}}

\providecommand\dedic{\thanks{Supported by
grant IAA100190902 of GA AV \v CR, Center of Excellence CE-ITI under the grant
P202/12/G061 of GA \v CR, and RVO: 67985840.}}
\author{Emil Je\v r\'abek\dedic\\[\medskipamount]
Institute of Mathematics of the Academy of Sciences\\
\small \v Zitn\'a 25,
115\:67 Praha 1,
Czech Republic,
email: \texttt{jerabek@math.cas.cz}
}

\def\lh#1{\lVert#1\rVert}
\def\pmi{\phantom-}

\linenumbersfalse

\def\dedic{\thanks{Supported by
grant IAA100190902 of GA AV \v CR, Center of Excellence CE-ITI under the grant
P202/12/G061 of GA \v CR, and RVO: 67985840. Part of the research was done
while visiting the Isaac Newton Institute in Cambridge.}}

\title{Integer factoring and modular square roots}

\begin{document}
\maketitle

\begin{abstract}
Buresh-Oppenheim proved that the $\np$ search problem to find
nontrivial factors of integers of a special form belongs to
Papadimitriou's class~$\ppa$, and is probabilistically reducible to a
problem in~$\ppp$. In this paper, we use ideas from bounded arithmetic
to extend these results to arbitrary integers. We show that general
integer factoring is reducible in randomized polynomial time to a
$\ppa$ problem and to the problem $\wpig\in\ppp$. Both reductions can
be derandomized under the assumption of the generalized Riemann
hypothesis. We also show (unconditionally) that $\ppa$ contains some
related problems, such as square root computation modulo~$n$, and
finding quadratic nonresidues modulo~$n$.
\end{abstract}

\section{Introduction}

Integer factoring is one of the best-known problems in complexity
theory which is in $\np$, but is not known to be polynomial-time
computable. In particular, the assumed hardness of factoring has
various applications in cryptography. Papadimitriou~\cite{papa:parity}
introduced several classes of search problems based on parity
arguments and related combinatorial principles. He showed that many
natural search problems from diverse areas of mathematics belong to
one of these classes, and he posed as an open problem whether the same
holds for integer factoring.

The first step to answer Papadimitriou's question was undertaken by
Buresh\hyph Oppenheim~\cite{jbo}. He proved that factoring of ``good''
integers (odd integers~$n$ such that $-1$ is not a quadratic residue
modulo~$n$) such that $n\equiv1\pod4$ belongs to the search class $\ppa$,
and factoring of good integers is probabilistically poly-time
reducible to a $\ppp$ problem. (Note that an odd integer is good iff it
has a prime divisor $p\equiv-1\pod4$.)

The purpose of this paper is to exhibit similar
reductions for factoring of arbitrary integers. We show
that factoring is probabilistically poly-time reducible to a $\ppa$
problem, as well as to $\wpig$, which is a $\ppp$ problem. (A similar
probabilistic reduction of factoring to~$\ppp$ was also independently
found by Buresh-Oppenheim~\cite{jbo:comm}.) We isolate
a convenient intermediate problem, which we call $\fsr$: given
integers $n$ and~$a$ such that the Jacobi symbol $\tljk an=1$, find
either a proper divisor of~$n$, or a square root of~$a$ modulo~$n$. It
is not hard to show that factoring is probabilistically poly-time
reducible to $\fsr$.

The main technical ingredient of our work is to demonstrate that
$\fsr\in\ppa$. The high-level idea of the proof comes from bounded
arithmetic. Je\v r\'abek~\cite{ej:flt} introduced an arithmetical
theory $\sss+\coun_2(\pv)$ related to~$\ppa$, and established that this
theory can prove the quadratic reciprocity theorem and other
properties of the Jacobi symbol, which together imply the soundness of
the usual poly-time algorithm for the Jacobi symbol. In particular,
$\sss+\coun_2(\pv)$ proves the totality of $\fsr$, and then an
application of a garden-variety witnessing theorem yields $\fsr\in\ppa$.
However, since this paper is intended for a general computational
complexity audience, we include a self-contained direct proof of this
result, we do not assume any prior knowledge (or posterior, for that
matter) of bounded arithmetic on the part of the reader.

All probabilistic reductions in this paper can be derandomized if we
assume the generalized Riemann hypothesis ($\grh$). In particular,
$\grh$ implies that factoring is in $\ppa\cap\ppp$ (and moreover, it
is poly-time reducible to $\wpig$). We also show unconditionally that
several problems concerning quadratic residues have deterministic
Turing reductions to $\fsr$, and as such are in~$\ppa$: for one, given
$n$ and~$a$, we can find either a square root of~$a$ modulo~$n$, or a
suitable witness that $a$ is a quadratic nonresidue. For another,
given an odd~$n$
which is not a perfect square, we can find an~$a$ such that $\tljk
an=-1$ (in particular, $a$ is a quadratic nonresidue modulo~$n$).

The paper is organized as follows. In Section~\ref{sec:preliminaries},
we review basic concepts used in the paper to fix the notation.
Section~\ref{sec:search-compl-fact} presents our main results, except
for the somewhat complex proof of $\fsr\in\ppa$, which is given
separately in Section~\ref{sec:fsrinppa}. Some concluding remarks
follow in Section~\ref{sec:conclusion}.

\section{Preliminaries}\label{sec:preliminaries}

An \emph{$\np$ search problem} is given by a poly-time computable
relation~$R(x,y)$ such that $R(x,y)$ implies $\lh y\le\lh x^c$ for
some constant~$c$, the problem is to find a~$y$ satisfying~$R(x,y)$
given~$x$. (We use $\lh x$ to denote the length of~$x$; most
of our algorithms work with integers, and we reserve $\abs x$
for the absolute value of~$x$. We also warn the reader that we will
often call our binary integers~$n$, we will not use the convention
that $n$ implicitly denotes the length of the input.) For brevity, we
may use $R$ to denote the search problem itself. A search problem~$R$
is \emph{total} if for every~$x$ there exists a~$y$ such that
$R(x,y)$. Unless indicated otherwise, all search problems below will
be assumed to be total $\np$ search problems.

We will often specify $\np$ search problems in the form
``given an~$x$ such that~$P(x)$, find a~$y$ satisfying~$R(x,y)$'',
where $P$ is a poly-time condition. In order to make it formally a
total search problem, this formulation will be understood to denote
the problem associated with the relation $(\neg P(x)\land
y=0)\lor(P(x)\land R(x,y))$.

A search problem~$R$ is \emph{many-one reducible} to a search
problem~$S$, written as~$R\le_mS$, if there are poly-time
functions~$f,g$ such
that $S(f(x),y)$ implies $R(x,g(x,y))$. $R$ is \emph{Turing-reducible}
to~$S$, written as~$R\le_TS$, if there exists a poly-time oracle Turing
machine~$M$ (where the oracle returns strings rather than yes/no answers) such that on input~$x$,
$M$~computes a~$y$ solving~$R(x,y)$ whenever all answers of the oracle
are correct solutions of~$S$. The class of all search problems~$R$
such that $R\le_TS$ will be denoted $\fp^S$.
If $C$ is a class of search problems, we write $R\le_mC$ if $R\le_mS$
for some~$S\in C$, and similarly for $R\le_TC$, $\fp^C$, as well as
other reduction notions mentioned below.

Let a circuit $C\colon\two^n\to\two^n$ (here, $\two=\{0,1\}$) encode an
undirected graph $G=\p{V,E}$, where $V=\two^n\bez\{0^n\}$, and
$\{u,v\}\in E$ iff $u,v\in V$, $u\ne v$, $C(u)=v$, and $C(v)=u$.
Notice that $G$ is a partial matching. $\lone$ is the following search
problem: given~$C$, find $u\in V$ unmatched by~$G$. The class $\ppa$
(for ``polynomial parity argument'') consists of
all search problems many-one reducible to $\lone$. (This
is not Papadimitriou's definition of~$\ppa$, it comes
from~\cite{bceip}, where it is shown to be equivalent to the original
one.) By abuse of notation, we will also use $\lone$ to denote the
following variant of the problem. Let $f(a,x)$,~$g(a)$ be poly-time
functions such that for every~$a$, $g(a)$ is an odd natural number,
and the function $f_a(x):=f(a,x)$ is an involution (i.e., $f_a(f_a(x))=x$) on the integer
interval $[0,g(a))$. Then the problem is, given~$a$ to find an~$x<g(a)$
which is a fixpoint of~$f_a$ (i.e., $f_a(x)=x$).
We will often use the fact that $\ppa$ is closed under Turing
reductions:
\begin{Thm}[Buss and Johnson \cite{buss-john}]\th\label{thm:bj}
$\fp^\ppa=\ppa$.\noproof
\end{Thm}

The class $\ppp$ (for ``polynomial pigeonhole principle'') consists of problems many-one reducible to~$\pig$,
which is the following problem: given a circuit
$C\colon\two^n\to\two^n$, find either a pair $u\ne v$ such that
$C(u)=C(v)$, or a~$u$ such that $C(u)=0^n$. If $p(n)$ is any polynomial
such that $p(n)>n$ for every~$n$, let $\wpig^{2^{p(n)}}_{2^n}$ denote
the following problem: given a circuit $C\colon\two^{p(n)}\to\two^n$,
find $u\ne v$ such that $C(u)=C(v)$. We define
$\wpig:=\wpig^{2^{n+1}}_{2^n}$; the choice of~$n+1$ here does not
matter:
\begin{Lem}\th\label{lem:wpig}
For any polynomial $p$ as above,
$\wpig\equiv_m\wpig^{2^{p(n)}}_{2^n}$.
\end{Lem}
\begin{Pf}
Given a circuit $C(\vec x,u)\colon\two^n\times\two\to\two^n$, we put
$m=p(n)-n$, and we construct a circuit
$D\colon\two^n\times\two^m\to\two^n$ by
$D(\vec x,u_0,\dots,u_{m-1})=C(\cdots(C(C(\vec x),u_0),u_1)\dots,u_{m-1})$.
Given $\p{\vec x,\vec u}\ne\p{\vec x',\vec u'}$ such that
$D(\vec x,\vec u)=D(\vec x',\vec u')$, we find the largest $i<m$ such
that $\p{\vec y,u_i}\ne\p{\vec y',u_i'}$, where
$\vec y^{(\prime)}=C(\cdots(C(C(\vec
x^{(\prime)}),u^{(\prime)}_0),u^{(\prime)}_1)\dots,u^{(\prime)}_{i-1})$.
Then $C(\vec y,u_i)=C(\vec y',u'_i)$.
\end{Pf}
The class of all search problems many-one reducible to~$\wpig$ does
not seem to have an established name in the literature, although it
clearly deserves one. In analogy with~$\ppp$, we can call it $\pwpp$
for ``polynomial weak pigeonhole principle''. Note that neither $\ppp$
nor~$\pwpp$ is known to be closed under Turing reductions. The proof
of \th\ref{lem:wpig} also implies that problems of the following kind
belong to~$\pwpp$; we will denote them all as $\wpig$ by abuse of
notation. Let $\ep>0$ be a constant, and $f,g$ poly-time function such
that for any $a$, $g(a)>0$, and~$f_a(x):=f(a,x)$ maps the interval
$\bigl[0,\cl{(1+\ep)g(a)}\bigr)$ into $\bigl[0,g(a)\bigr)$. Then the problem is, given
$a$, to find $u<v<\cl{(1+\ep)g(a)}$ such that $f_a(u)=f_a(v)$.

Apart from $\le_m$ and~$\le_T$, we will also need randomized
reductions. We will use several different versions to be able to
state our results precisely; the definitions below are not standard,
but we believe they are quite natural.

For any constant $0<\ep<1$, we say that $R$ is
\emph{probabilistically many-one reducible to~$S$ with error~$\ep$},
written as~$R\le_m^{\rp,\ep}S$, if there is a
polynomial~$p$ and poly-time functions $f(x,r)$ and~$g(x,r,y)$ such
that for every~$x$,
$$\Pr_{\lh r=p(\lh x)}[\forall y\,[S(f(x,r),y)\TO R(x,g(x,r,y))]]\ge1-\ep.$$
We say that $R$ is
\emph{probabilistically many-one reducible to~$S$ with controlled
error}, written as~$R\le_m^\rp S$, if there is a polynomial~$p$
and poly-time functions $f(x,1^k,r)$ and~$g(x,1^k,r,y)$ such that for
every $x$ and~$k$,
$$\Pr_{\lh r=p(\lh x,k)}[\forall y\,[S(f(x,1^k,r),y)\TO R(x,g(x,1^k,r,y))]]\ge1-2^{-k}.$$
$R$ is \emph{probabilistically Turing-reducible to~$S$}, written
as~$R\le_T^\rp S$, if there exists a polynomial~$p$ and a poly-time
oracle Turing machine~$M$ such that
$$\Pr_{\lh r=p(\lh x)}[\text{every sound run of $M(x,r)$ solves $R(x,y)$}]\ge1/2,$$
where a run is sound if all oracle answers are correct solutions
of~$S$. Note that the constant~$1/2$ here is arbitrary, as we can
decrease the error from any constant~$\ep>0$ to any other
constant (or to controlled error as above) in the usual way: we
can check solutions of~$R$, hence we can run the machine several times
with independent choices of~$r$, and return the first correct solution
to the search problem. We denote by $\tfrp^S$ the class of all~$R$
such that $R\le_T^\rp S$. We observe that we can split a randomized
Turing reduction as a randomized many-one reduction followed by a
deterministic Turing reduction; this is particularly useful when $S$
is from a Turing-closed class such as $\ppa$.
\begin{Lem}\th\label{lem:rptur}
$\tfrp^S\le_m^\rp\fp^S$.
\end{Lem}
\begin{Pf}
Assume that $R\le_T^\rp S$ and $M^S$ is the Turing machine from the
definition.
Let $T$ be the following search problem: given $x$ and~$r$, find a
sound run of $M^S(x,r)$. It is easy to see that $T$ is a total $\np$ search
problem, and $R\le_m^\rp T\le_T S$.
\end{Pf}
\begin{Lem}\th\label{lem:rptiter}
$\tfrp^{\tfrp^S}=\tfrp^S$.
\end{Lem}
\begin{Pf}
In view of \th\ref{lem:rptur} and the obvious transitivity of
$\le_m^\rp$, it suffices to show that $\tfrp^S$ is closed under
deterministic Turing reductions. Let thus $T\in\tfrp^S$, and $M^T$ be
a poly-time oracle machine solving $R(x,y)$. Since answers of the
oracle have polynomial length, the total number of sound runs of~$M$
on input~$x$ is bounded by $2^{\lh x^c}$ for some constant~$c$. Using
the above-mentioned amplification of success rate, we can find a randomized poly-time
machine $N^S$ solving $T$ with error $2^{-\lh x^{c+1}}$. If we
then use $N$ to answer $M$'s oracle queries while reusing the same
pool of random bits for every call, all but a fraction of $2^{\lh x^c}2^{-\lh
x^{c+1}}\ll1$ of the random choices will be good for every possible
run of the combined machine.
\end{Pf}
A many-one reduction of~$R$ to~$S$ is supposed to construct a valid
instance of~$S$ from whose solution it can recover a solution to the
original problem. In the case of~$\le_m^\rp$, the reduction algorithm
succeeds in doing this only with some bounded probability. 
It will be also useful to consider stronger notions of reduction where
we can check before consulting the oracle whether the particular
choice of random bits leads to the desired result. The reduction
function may abandon the computation with some bounded probability,
but if it does not, then any valid solution of~$S$ gives a solution
of~$R$. Alternatively, we could repeat the computation until we find a
``good'' instance of~$S$, and only then pass the query to the oracle;
in this way, the reduction always succeeds, but only its expected
running time is polynomial.

Formally,
$R$ is \emph{probabilistically zero-error many-one reducible to~$S$},
written as~$R\le_m^\zpp S$, if there is a polynomial~$p$,
poly-time functions $f(x,r)$ and~$g(x,r,y)$, and a poly-time predicate
$h(x,r)$, such that
\begin{enumerate}
\item\label{item:prob} $\Pr_{\lh r=p(\lh x)}[h(x,r)]\ge1/2$,
\item if $h(x,r)$ and $S(f(x,r),y)$, then $R(x,g(x,r,y))$.
\end{enumerate}
Similarly, $R$ is \emph{probabilistically zero-error Turing-reducible to~$S$},
written as~$R\le_T^\zpp S$, if there is a polynomial~$p$, a poly-time predicate
$h(x,r)$, and a poly-time
oracle Turing machine~$M$, such that~\eqref{item:prob}, and
if~$h(x,r)$, then every sound run of~$M^S(x,r)$ solves~$R(x,y)$.
Again, the constant~$1/2$ is arbitrary, we can amplify the success
rate from any constant~$\ep>0$ to~$1-2^{-k}$ (even for many-one
reductions). Let $\tfzpp^S$ denote the class of all problems~$R$ such
that $R\le_T^\zpp S$. Note that if there is no oracle, $\tfzpp=\tfrp$.

$\fac$ is the following search problem: given a composite integer~$n$, find a
nontrivial divisor of~$n$. We define $\ffac$ to be the following
problem: given an integer~$n>0$, find a
sequence $\p{p_i:i<k}$ of primes such that $n=\prod_{i<k}p_i$ (here
and below, the empty product is defined to be~$1$). Note that $\fac$ and
$\ffac$ are total $\np$ search problems as primality testing is
poly-time (Agrawal, Kayal, and Saxena \cite{aks:prime}).
Clearly, $\fac\le_m\ffac\le_T\fac$.

We will denote the divisibility relation by $d\mid n$, modular
congruences by $a\equiv b\pod n$, and greatest common divisors by
$(a,b)$. An integer~$a$ is a \emph{quadratic residue}
modulo~$n$ if $a\equiv b^2\pod n$ for some~$b$. The \emph{Legendre
symbol} is defined for any integer~$a$ and an odd prime~$p$ by
$$\ljk ap=\begin{cases}
\pmi0&p\mid a,\\
\pmi1&p\nmid a\text{ and $a$ is a quadratic residue mod $p$,}\\
-1&p\nmid a\text{ and $a$ is a quadratic nonresidue mod $p$.}
\end{cases}$$
More generally, the \emph{Jacobi symbol} is defined for any odd~$n>0$
by
$$\ljk an=\prod_{i<k}\ljk a{p_i},$$
where $n=\prod_{i<k}p_i$ is the prime factorization of~$n$. We will also
write $\tljk an$ instead of $\ljk an$ for typographical convenience.
\begin{figure}
\begin{algo}
$r\ass 1$
\key{while} $a\ne0$ \key{do}:
	\key{if} $a<0$ \key{then}:
		$a\ass-a$
		$r\ass-r$ \key{if} $n\equiv-1\pod 4$
	\key{while} $a$ is even \key{do}:
		$a\ass a/2$
		$r\ass-r$ \key{if} $n\equiv\pm 3\pod 8$
	swap $a$ and $n$
	$r\ass-r$ \key{if} $a\equiv n\equiv-1\pod 4$
	reduce $a$ modulo $n$ so that $\abs a<n/2$
\key{if} $n>1$ \key{then} output $0$ \key{else} output $r$
\end{algo}
\caption{An algorithm for the Jacobi symbol $\tljk an$}
\label{fig:jac}
\end{figure}
A \emph{Dirichlet character} of modulus~$n$ is a group
homomorphism $\chi\colon(\Z/n\Z)^*\to\C^*$. A character is
\emph{principal} if it only assumes the value~$1$, and
\emph{real} if it takes values in $\{1,-1\}$. Characters can be
lifted to mappings $\Z\to\C$ by putting $\chi(a)=0$ when
$(a,n)\ne1$. Note that for any odd positive~$n$, $\chi_n(x)=\tljk xn$ is a
real character of modulus~$n$ (in particular, $\tljk
an\tljk bn=\tljk{ab}n$), which is principal iff $n$ is a perfect
square. The characters $\chi_n$ are called \emph{quadratic}. The
\emph{quadratic reciprocity theorem} states that
for any coprime odd $n,m>0$,
$$\ljk nm\ljk mn=\begin{cases}
-1&\text{if }n\equiv m\equiv-1\pod4\\
\pmi1&\text{otherwise.}
\end{cases}$$
Together with the supplementary laws
$$\ljk{-1}n=\begin{cases}\pmi1&n\equiv\pmi1\pod4\\-1&n\equiv-1\pod4\end{cases}
\qquad\ljk2n=\begin{cases}\pmi1&n\equiv\pm1\pod8\\-1&n\equiv\pm3\pod8\end{cases}$$
it implies that the Jacobi symbol is poly-time computable
(see Figure~\ref{fig:jac}).

The \emph{generalized Riemann hypothesis\footnote{Also called the
extended Riemann hypothesis ($\M{ERH}$). The nomenclature of various
extensions of $\M{RH}$ varies wildly in the literature. We chose
to denote the $\M{RH}$ for Dirichlet $L$-functions by $\grh$ as this
name seems to be more specific, whereas $\M{ERH}$ is often used for
other generalizations of $\M{RH}$, such as the $\M{RH}$ for Dedekind
$\zeta$-functions, or $L$-functions of Hecke characters.} 
($\grh$)} states that for every
Dirichlet character~$\chi$, all zeros of its associated $L$-function
$L(\chi,s)$ in the critical strip $0<\Re(s)<1$ satisfy $\Re(s)=1/2$. Let
$\grhq$ denote the special case of $\grh$ for quadratic characters
$\chi$. We will use the following result of Bach~\cite{bach}, refining the work of Ankeny~\cite{ank}.
\begin{Thm}\th\label{thm:bach}
Assume $\grhq$. If\/ $\chi$ is a nonprincipal quadratic
character with modulus~$n$, there exists $0<a<2(\ln n)^2$ such that\/
$\chi(a)\ne1$.
\noproof\end{Thm}

\section{Search complexity of factoring}\label{sec:search-compl-fact}

In this section, we are going to describe our main result
(\th\ref{thm:fac}) on the
relationship of factoring to the classes $\ppa$ and $\ppp$ ($\pwpp$).
Rather than working directly with $\fac$, it will be convenient to
consider other related problems.
\begin{Def}
Let $\fsr$ denote the following problem: given an odd integer~$n>0$
and an integer~$a$ such that $\tljk an=1$, find either a nontrivial
divisor of~$n$, or a square root of~$a$ modulo~$n$.

We also give names to some special cases of $\fsr$. $\fsrm$ denotes
the problem, given odd~$n>0$ and integers $a$ and~$b$, to find a
nontrivial divisor of~$n$ or a square root of one of~$a$, $b$, or~$ab$
modulo~$n$.

$\wfsr$ is the following problem: given an odd~$n>0$ and $a$,~$b$ such
that $\tljk an=1$ and $\tljk bn=-1$, find a nontrivial divisor of~$n$,
or a square root of~$a$ modulo~$n$.
\end{Def}
We start with basic dependencies between these problems.
\begin{Lem}\th\label{lem:easy}
\ \begin{enumerate}
\item\label{item:fsrm} $\wfsr\le_m\fsrm\le_m\fsr$;
\item\label{item:fac} $\wfsr\le_m\fac$.
\end{enumerate}
\end{Lem}
\begin{Pf}
\eqref{item:fsrm}: $\wfsr$ is a special case of $\fsrm$, since
$\tljk an=1$ and $\tljk bn=-1$ imply that neither~$b$ nor~$ab$ is
a quadratic residue modulo~$n$. Given an instance of $\fsrm$, the
multiplicativity of the Jacobi symbol implies that $\tljk xn=1$ for
some $x\in\{a,b,ab\}$. We can choose such an~$x$ as the Jacobi symbol is
poly-time computable, and then we pass it to $\fsr$.

\eqref{item:fac}: If $n$ is prime, we can compute a square root of~$a$
modulo~$n$ in polynomial time using the Shanks--Tonelli algorithm.
This algorithm is deterministic if we provide it with a quadratic
nonresidue, which we can:~$b$. If $n$ is composite, we pass it to
$\fac$.
\end{Pf}
\begin{Lem}\th\label{lem:rand}
\ \begin{enumerate}
\item\label{item:randfsr} $\fsr\le_m^\zpp\wfsr$;
\item\label{item:randfac} $\fac\le_m^{\rp,1/2}\fsr$;
\item\label{item:randwfsr} $\fac\le_m^{\rp,1/2}\wfsr$.
%\item\label{item:randffac} $\ffac\le_T^\rp\wfsr$.
\end{enumerate}
\end{Lem}
\begin{Pf}
\eqref{item:randfsr}: If $n$ is a perfect square, we can return~$\sqrt
n$ as its nontrivial divisor (unless it is $1$, in which case we can
return~$0$ as the square root of~$a$). Otherwise $\chi_n$ is a
nonprincipal real character, hence with probability at least $1/2$,
a randomly chosen $0<b<n$ either shares a factor with~$n$ (in which case
we can return $(n,b)$ as a nontrivial divisor) or satisfies $\tljk
bn=-1$, and we can pass it to $\wfsr$.

\eqref{item:randfac}: If $n$ is even or a perfect power, we can factor
it directly, hence we may assume $n$ is odd and it has $k\ge2$
distinct prime divisors. We consider the following reduction. We choose a
random $0<a<n$. If $(a,n)\ne1$, we can return it as a nontrivial
divisor of~$n$, otherwise we pass $n,a$ to a $\fsr$ oracle.

Since $\chi_n$ is a nonprincipal real character, we have
$\tljk an=1$ for a half of all residues from $(\Z/n\Z)^*$.
On the other hand, if $n=\prod_{i<k}p_i^{e_i}$, where the~$p_i$ are distinct
primes, then $a$ coprime to~$n$ is a quadratic residue modulo~$n$ iff
$\tljk a{p_i}=1$
for every~$i<k$. Using the Chinese remainder theorem, a fraction
$2^{-k}$ of $(\Z/n\Z)^*$ are quadratic residues. Thus, with
probability at least $1/2-2^{-k}\ge1/4$, the chosen~$a$ either shares
a factor with~$n$, or it satisfies
$\tljk an=1$ while not being a quadratic residue, hence the $\fsr$
oracle must give us a factor of~$n$.

We can amplify the success probability to~$1/2$ by observing that
residues~$a$ such that $\tljk an=1$ are poly-time samplable. We assume
w.l.o.g.\ that $n$ is not a perfect square. The
reduction works as follows. We choose random $0<a,b<n$. If $(n,a)\ne1$
or~$(n,b)\ne1$, we can factorize~$n$. Otherwise, we let $c$ be the first
residue from the list~$a,b,ab$ which satisfies $\tljk cn=1$, and we
call $\fsr(n,c)$. It is easy to see that the induced distribution
of~$c$ is the uniform distribution over $\{c<n:\tljk cn=1\}$, hence
conditioned on $(a,n)=(b,n)=1$, $c$~is a quadratic nonresidue with
probability $1-2^{1-k}\ge1/2$.

\eqref{item:randwfsr}: $\fsr\le_m^\rp\wfsr$ by~\eqref{item:randfsr}
and amplification of the success rate of~$\le_m^\zpp$, hence
$\fac\le_m^{\rp,1/2+\ep}\wfsr$ for any~$\ep>0$ by~\eqref{item:randfac}.
We can get rid of the~$\ep$ by observing that the proof
of~\eqref{item:randfac} actually shows $\fac\le_m^{\rp,1/2-1/\sqrt
n}\fsr$, taking into account residues that share a factor with~$n$. We
can reduce the error of the~$\le_m^\zpp$ reduction in
\eqref{item:randfsr} to $1/\sqrt n$, hence
$\fac\le_m^{\rp,1/2}\wfsr$.
%
%\eqref{item:randffac} follows from \eqref{item:randfsr},
%\eqref{item:randfac}, and \th\ref{lem:rptiter}.
\end{Pf}

We remark that there is another well-known randomized reduction of factoring to
square root computation modulo~$n$ due to Rabin~\cite{rab:cryp}, but it is suited for a different model. In the
notation above, the basic idea of Rabin's reduction is that we choose a random $1<a<n$,
and if it is coprime to~$n$, we pass $n,a^2$ to the $\fsr$ oracle. If the
oracle were implemented as a (deterministic or randomized) algorithm working independently of the reduction without
access to its random coin tosses,
we would have a $1/2$ chance that the root $b$ of~$a^2$ returned by
the oracle satisfies $a\nequiv\pm b\pod n$, allowing us to factorize~$n$.
However, this does not work in our setup. According to the
definition of a search problem reduction, the reduction function must be
able to cope with \emph{any} valid answer to the oracle query---there
is no implied guarantee that oracle answers are computed
independently of the environment. In particular, it may happen
the oracle is devious enough to always return the root $b=a$ we
already know.

What we need now is to show that $\fsr$ or some of its variants
belongs to $\ppa$ and $\pwpp$.
\begin{Thm}\th\label{thm:fsr}
$\fsr\in\ppa$.
\end{Thm}
We will prove \th\ref{thm:fsr} in the next section, as the argument is
a bit involved.

For the pigeonhole principle, we have the following reduction, whose
idea comes from the proof of the multiplicativity of the Legendre
symbol in $\idz+\wphp(\delz)$ by Berarducci and
Intrigila~\cite{berint}.
\begin{Thm}\th\label{thm:wphp}
$\fsrm\in\pwpp$.
\end{Thm}
\begin{Pf}
Assume we are given an odd~$n>1$, and integers $a$,~$b$. If $a$ or~$b$
shares a factor with~$n$, we can return $(n,a)$ or $(n,b)$, resp., as
a nontrivial divisor of~$n$, we thus assume both are coprime to~$n$.
Consider the following poly-time function
$f\colon\{0,1,2\}\times[1,(n-1)/2]\to[1,n-1]$:
$$f(i,x)=\begin{cases}a_ix^2\bmod n&\text{ if $(n,x)=1$,}\\
x&\text{otherwise,}
\end{cases}$$
where $a_0=1$, $a_1=a$, $a_2=b$. Since the domain of~$f$ is
$3/2$~times larger than its range, we can use $\wpig$ to find a collision
$f(i,x)=f(j,y)$, $\p{i,x}\ne\p{j,y}$. We may assume $(n,x)=(n,y)=1$,
as otherwise we can factor~$n$. If $i=j$, then $x^2\equiv y^2\pod n$,
but $x\nequiv\pm y\pod n$, hence $(n,x-y)$ is a nontrivial divisor
of~$n$. If $i<j$, then $a_ja_i^{-1}\equiv(xy^{-1})^2\pod n$ (where the
inverses are also modulo~$n$), hence $xy^{-1}$ is a square root
of~$a$, $b$, or~$ba^{-1}$ modulo~$n$. In the last case, $axy^{-1}$ is
a square root of~$ab$.
\end{Pf}
We mention that essentially the same reduction of\/~$\fac$ to~$\wpig$
by means of\/~$\fsrm$ was used in a different context
in~\cite[Thms.~4.1--2]{ej:wphpvar}, and a similar reduction was
independently discovered by Buresh-Oppenheim~\cite{jbo:comm}.

While we do not know whether $\pwpp$ is closed under general Turing
reductions, the next lemma shows that it is closed under
\emph{nonadaptive} Turing reductions.
\begin{Lem}\th\label{lem:para}
The following problem, denoted $\wpig^\|$, is in $\pwpp$: given a
sequence $\p{C_i:i<m}$ of circuits
$C_i\colon\two^{n_i+1}\to\two^{n_i}$, find sequences $\p{u_i:i<m}$ and
$\p{v_i:i<m}$ such that $u_i,v_i\in\two^{n_i}$, $u_i\ne v_i$, and
$C_i(u_i)=C_i(v_i)$ for each~$i<m$.
\end{Lem}
\begin{Pf}
Put $n=\max_in_i$. We can pad each~$C_i$ to $n$ output bits by
considering the circuit
$C'_i\colon\two^{n-n_i}\times\two^{n_i+1}\to\two^{n-n_i}\times\two^{n_i}$
defined by $C'_i(x,u)=\p{x,C_i(u)}$, hence we may assume $n=n_i$
without loss of generality. By~\th\ref{lem:wpig}, we can amplify each
$C_i$ to a circuit $D_i\colon\two^{mn+1}\to\two^n$, and we
define a circuit $D\colon\two^{mn+1}\to(\two^n)^m$ by
$D(u)=\p{D_i(u):i<m}$. Using a call to~$\wpig$, we find $u\ne v$ such
that~$D(u)=D(v)$. Then $D_i(u)=D_i(v)$ for each~$i$, and we can
compute $u_i\ne v_i$ such that~$C_i(u_i)=C_i(v_i)$.
\end{Pf}

We obtain the main result of this paper by putting everything together:
\begin{Thm}\th\label{thm:fac}
\ \begin{enumerate}
\item\label{item:facppa} $\fac,\ffac\le_m^\rp\ppa$;
\item\label{item:facpwpp} $\fac\le_m^\rp\pwpp\sset\ppp$ and
$\ffac\le_m^\rp\fp^\pwpp\sset\fp^\ppp$.
\end{enumerate}
\end{Thm}
\begin{Pf}
\eqref{item:facppa}: $\ffac$ is in~$\tfrp^\fsr$ by
\th\ref{lem:rand,lem:rptiter}, hence in~$\tfrp^\ppa$
by~\th\ref{thm:fsr}. This implies
$\ffac\le_m^\rp\fp^\ppa=\ppa$ by \th\ref{lem:rptur,thm:bj}.

\eqref{item:facpwpp}:
We have $\fac\le_m^{\rp,1/2}\pwpp$ by \th\ref{lem:rand,thm:wphp}.
Given $k$ in unary, we can reduce the error to~$2^{-k}$ with $k$
parallel calls to a $\wpig$ oracle, which implies
$\fac\le_m^\rp\wpig^\|\in\pwpp$ by~\th\ref{lem:para}.
As in~\eqref{item:facppa}, we have $\ffac\le_T^\rp\pwpp$,
hence $\ffac\le_m^\rp\fp^\pwpp$ by~\th\ref{lem:rptur}.
\end{Pf}
It would be desirable to derandomize the results in
\th\ref{thm:fac}. We are only able to do it under an extra assumption.

\pagebreak[2]
\begin{Thm}\th\label{thm:grh}
Assume $\grhq$.
\begin{enumerate}
\item $\fac\equiv_m\fsr\equiv_m\wfsr\equiv_m\fsrm$;
\item $\fac,\ffac\in\ppa$;
\item $\fac\in\pwpp$, $\ffac\in\fp^\pwpp$.
\end{enumerate}
\end{Thm}
\begin{Pf}
It suffices to derandomize the reductions in
\th\ref{lem:rand}~(\ref{item:randfsr},\ref{item:randfac}). For
$\fsr\le_m\wfsr$, note that \th\ref{thm:bach} guarantees that we can
find a suitable $b<2(\ln n)^2=O(\lh n^2)$.

For $\fac\le_m\fsr$, it suffices to show that for any odd~$n$ which is not a
prime power, there exists an $0<a<(\ln n)^{O(1)}$ such that either
$(a,n)>1$, or $\tljk an=1$ and $a$ is a quadratic nonresidue
modulo~$n$; the latter means that $\tljk ap=-1$ for some prime $p\mid
n$.

We can assume that $(a,n)=1$ for every $0<a<2(\ln n)^2$, otherwise we
are done. Let $p$ be a prime divisor of~$n$ such that, if possible,
the exponent of~$p$ in the prime factorization of~$n$ is even, so that
$n/p$ is not a perfect square. Then 
$\chi_{n/p}$ is a nonprincipal quadratic character, and there is
$0<u<2(\ln(n/p))^2$ such that $\tljk u{n/p}=-1$ by \th\ref{thm:bach}.
This implies $\tljk un=-\tljk up$. If $\tljk un=1$, we can take
$a=u$. Otherwise, we have $\tljk un=-1$ and $\tljk up=1$. Since
$\chi_p$ is also a nonprincipal quadratic character, there
is $0<v<2(\ln p)^2$ such that $\tljk vp=-1$. If $\tljk vn=1$, we can
take $a=v$, otherwise we take $a=uv$. Either way,
$a<4(\ln p)^2(\ln(n/p))^2<\frac14(\ln n)^4$.
\end{Pf}

We can use $\fsr$ with constant~$a$ to obtain special cases of
factoring that are unconditionally in deterministic $\ppa$, see
\th\ref{exm:fsra}. In fact, we can factor~$n$ as long as there exists
a quadratic nonresidue $a=(\log n)^{O(1)}$ such that $\tljk an=1$. We
can express this more perspicuously as follows.
\begin{Def}\th\label{def:strong}
Let $s>0$. An integer~$n$ is \emph{$s$-strongly composite,} if
we can write $n=n_0n_1$ so that neither $n_0$ nor~$n_1$
is a quadratic residue modulo~$s$.
\end{Def}
Notice that an odd integer is 4good in the sense of~\cite{jbo} iff
it is $4$-strongly composite.
\begin{Thm}\th\label{thm:strcomp}
For any constant~$c$, the following problem is in~$\ppa$: given
an~$n>0$ which is $s$-strongly composite for $s=\fl{(\log n)^c}!$, find a
nontrivial divisor of~$n$.
\end{Thm}
\begin{Pf}
We can assume w.l.o.g.\ that $n$ is coprime to $\fl{(\log n)^c}!$
(hence odd).
It suffices to show that there exists an~$a$ with $\abs a\le(\log n)^{2c}$ such
that $\tljk a{n_0}=\tljk a{n_1}=-1$.
Since $n_i$ is a quadratic nonresidue modulo~$s$, it is also a
quadratic nonresidue modulo~$s_i$, where $s_i=8$, or $s_i$ is an odd
prime divisor of~$s$, i.e., $s_i\le(\log n)^c$.

Assume first that both $n_0,n_1$ are quadratic nonresidues
modulo~$s_0$. If $s_0$ is odd, we put $a=s_0^*:=(-1)^{(s_0-1)/2}s_0$. Then
$\tljk a{n_i}=\tljk{n_i}{s_0}=-1$ by quadratic reciprocity. If
$s_0=8$, i.e., $n_0,n_1\nequiv1\pod8$, we choose $m\in\{3,5,7\}$ such
that $m\nequiv n_0,n_1\pod8$, and we put
$$a=\begin{cases}
-2&m=3,\\
-1&m=5,\\
2&m=7.
\end{cases}$$
Then $\tljk a{n_0}=\tljk a{n_1}=-1$.

If both $n_0,n_1$ are quadratic nonresidues modulo~$s_1$, we proceed
similarly.

Assume that $n_i$ is a quadratic residue modulo~$s_{1-i}$ for~$i=0,1$. Put
$$a_i=\begin{cases}
s_i^*&s_i\text{ is odd,}\\
-1&s_i=8,n_i\equiv3,7\pod8,\\
2&s_i=8,n_i\equiv5\pod8,
\end{cases}$$
and $a=a_0a_1$. Then $\tljk{a_i}{n_i}=-1$ and
$\tljk{a_{1-i}}{n_i}=1$, hence $\tljk a{n_i}=-1$.
\end{Pf}
Conversely, one can show that if $a$ is a quadratic nonresidue such
that $\tljk an=1$, then $n$ is $s$-strongly composite for any~$s$
divisible by $4a$.

In \th\ref{thm:strcomp}, we do not need $s$ to have the exact form
given there: it is only essential that the prime factorization of~$s$
is known.

It is not clear whether one can fully unconditionally derandomize
\th\ref{thm:fac}. While no deterministic polynomial-time algorithm
to find quadratic nonresidues is known without $\grh$, in $\ppa$ we
can do better:

\pagebreak[2]
\begin{Lem}\th\label{lem:nonres}
The following problem is in $\fp^\fsr\sset\ppa$: given an odd $n>1$,
find an~$a$ such that\/ $\tljk an=-1$, or a nontrivial divisor of~$n$.
\end{Lem}
\begin{Pf}
Consider the following algorithm. Put $a=-1$.
While $\tljk an=1$, repeat the following steps: call the $\fsr$
oracle; if it provides a factor of~$n$, we are done, otherwise we
replace $a$ with its square root modulo~$n$.

The algorithm must halt within $\log_2n$ iterations: if $a$ is a
$2^k$th root of~$-1$, its order in $(\Z/n\Z)^*$ is $2^{k+1}<n$.
\end{Pf}
Notice that, conversely, $\fsr$ is Turing-reducible to $\wfsr$
together with the problem from \th\ref{lem:nonres}.

In fact, $\fsr$ does the dual job of factoring and computing square
roots. In \th\ref{thm:fac} we have exploited its factoring capacity by
supplying it with quadratic nonresidues, but we can also use it the other
way round to obtain algorithms for finding square roots and
quadratic nonresidues modulo arbitrary integers. We start with the latter.
\begin{Thm}\th\label{prop:nonjac}
The following problem is in $\fp^\fsr\sset\ppa$: given an odd~$n$
which is not a perfect square, find
an~$a$ such that\/ $\tljk an=-1$.
\end{Thm}
\begin{Pf}
The algorithm maintains a sequence $\p{n_i:i<k}$ of integers~$n_i>1$
such that $n=\prod_{i<k}n_i$, and a sequence $\p{a_i:i<k}$, where some
of the~$a_i$ may be undefined, but if $a_i$ is defined, then
$\tljk{a_i}{n_i}=-1$. We initialize it with~$k=1$, $n_0=n$,
$a_0$~undefined, and we repeat in arbitrary order the following steps
until neither is applicable any more:
\begin{itemize}
\item If $n_i\ne n_j$ are such that $(n_i,n_j)>1$, we delete
$n_i$, $n_j$ from the sequence and replace them with $(n_i,n_j)$ (two
copies), $n_i/(n_i,n_j)$, and $n_j/(n_i,n_j)$, omitting those equal
to~$1$ (this can happen only for one of the four numbers, hence the
length of the sequence always increases). The $a_i$ entries
corresponding to the new numbers are undefined.
\item If $a_i$ is undefined, we call as an oracle the search problem
from \th\ref{lem:nonres} on~$n_i$. If it returns a nontrivial
divisor of~$n_i$, we expand the $n_j$ sequence as in the previous
step. Otherwise, it provides a value for~$a_i$.
\end{itemize}
Since $k\le\log n$, the algorithm must halt in $O(\lh n)$ steps.
When it does, all~$a_i$ are defined, and the $n_i$ entries are
pairwise equal or coprime,
hence we can write $n=\prod_{i\in I}n_i^{e_i}$ for some
$I\sset\{0,\dots,k-1\}$ and $e_i>0$, where $n_i$,~$i\in I$, are
pairwise coprime. Since $n$ is not a perfect square, we can pick $i\in
I$ such that $e_i$ is odd. By the Chinese remainder theorem, we can
compute an~$a$ such that $a\equiv a_i\pod{n_i^{e_i}}$ and
$a\equiv1\pod{n/n_i^{e_i}}$. Then
$$\ljk an=\prod_{j\in I}\ljk a{n_j}^{e_j}=(-1)^{e_i}=-1.\qedhere$$
\end{Pf}
\begin{Cor}\th\label{prop:nonres}
The following problem is in $\fp^\fsr\sset\ppa$: given~$n>2$, find
an~$a$ coprime to~$n$ which is a quadratic nonresidue modulo~$n$.
\end{Cor}
\begin{Pf}
If $n$ is a power of~$2$, we can return~$3$. Otherwise, we can write
$n=2^em^{2^k}$, where $m$ is odd and not a perfect square. By
\th\ref{prop:nonjac}, we can find $a$ such that $\tljk am=-1$. By
adding $m$ to~$a$ if necessary, we can make sure $a$ is odd, hence
$(n,a)=1$. Since $a$ is a quadratic nonresidue modulo~$m\mid n$, it is
also a nonresidue modulo~$n$.
\end{Pf}

Another problem we are going to reduce to $\fsr$ is the computation of
square roots modulo~$n$. A priori it is not clear how to formulate it
as a total $\np$ search problem, as the quadratic residuosity problem
is neither known nor assumed to be poly-time decidable. We can remedy this by
requiring the search problem to find something sensible also for
quadratic nonresidues.
\begin{Def}\th\label{def:sqwit}
Let $n$ be a positive integer. If $(a,n)=1$, a divisor $m\mid n$ is a
\emph{coprime nonsquare witness for~$a$ modulo~$n$} if
\begin{itemize}
\item $m$ is odd and $\ljk am=-1$, or
\item $m=4$ and $a\equiv3\pod4$, or
\item $m=8$ and $a\equiv5\pod 8$.
\end{itemize}
If $a$ is an arbitrary integer, an $m$ is a \emph{nonsquare
witness for~$a$ modulo~$n$}, if $m$ is not a perfect square, $m$ is
odd or~$2$, and there are $e$,~$b$, and~$j<e$ such
that $m^e\mid n$, $a=m^jb$, $(m,b)=1$,
and if $j$ is even, $m$ (if odd) or $4$ or~$8$ (if $m=2$) is a
coprime nonsquare witness for~$b$ modulo~$m^{e-j}$.

It is easy to see that the property of being a nonsquare witness is
poly-time decidable.

Let $\sroot$ denote the following search problem: given $n>0$ and~$a$,
find either a square root of~$a$ modulo~$n$, or a nonsquare witness
for~$a$ modulo~$n$.
\end{Def}

%\pagebreak[2]
\begin{Lem}\th\label{lem:sqwit}
If there exists a nonsquare witness for~$a$ modulo~$n$, then $a$ is a
quadratic nonresidue modulo~$n$.
\end{Lem}
\begin{Pf}
If $m$ is a coprime nonsquare witness for~$a$, then $a$ is a quadratic
nonresidue modulo~$m$, and a fortiori modulo~$n$.

Let $m$ be a nonsquare witness for~$a$, and let $e$,~$b$, and~$j$ be as in \th\ref{def:sqwit}. Assume for contradiction
$a\equiv(uc)^2\pod n$, where $(m,c)=1$, and $u\mid m^k$ for some~$k$. We
have $m^j\mid(uc)^2$, hence $m^j\mid u^2$. Moreover, if we write $u^2=m^jv$,
then $b\equiv vc^2\pod{m^{e-j}}$, hence $(m,v)=1$, i.e., $v=1$ and $m^j=u^2$.
Since $m$ is not a perfect square, this implies $j$ is even. However,
$b\equiv c^2\pod{m^{e-j}}$ contradicts the fact that $b$ has a
coprime nonsquare witness modulo~$m^{e-j}$.
\end{Pf}
Notice that $\sroot$ is a generalization of $\fsr$: a nonsquare
witness for~$a$ modulo~$n$ is a nontrivial divisor of~$n$, unless $n$
is odd and $\tljk an=-1$.

\begin{Thm}\th\label{thm:root}
$\sroot\in\fp^\fsr\sset\ppa$.
\end{Thm}
\begin{Pf}
Write $n=2^em$ with $m$ odd. In the first stage of our algorithm, we
keep a sequence $\p{n_i:i<k}$ of integers $n_i>1$ such that
$m=\prod_{i<k}n_i$, and a sequence of integers
$\p{u_i:i<k}$ where some~$u_i$ may be undefined. We maintain the
property that whenever $u_i$ is
defined, we can write $a=n_i^{j_i}a_i$ for some~$j_i$ so that
$(a_i,n_i)=1$, and we have $a_i\equiv u_i^2\pod{n_i}$. We start with
$k=1$, $n_0=m$ and $u_0$ undefined, and we repeat the
following steps until none of them are applicable any more:
\begin{itemize}
\item If $n_i\ne n_j$ are such that $(n_i,n_j)>1$, we delete
$n_i$, $n_j$ from the sequence and replace them with two copies of
$(n_i,n_j)$, $n_i/(n_i,n_j)$, and $n_j/(n_i,n_j)$ as in the proof of
\th\ref{prop:nonjac}.
\item If $n_i$ is a perfect square, we replace $n_i$ with two copies
of $\sqrt{n_i}$.
\item If $a=n_i^{j_i}a_i$ where $n_i\nmid a_i$, but $(n_i,a_i)>1$, we
replace $n_i$ with $(n_i,a_i)$ and $n_i/(n_i,a_i)$.
\item If $a=n_i^{j_i}a_i$ where $\tljk{a_i}{n_i}=1$, but $u_i$ is
undefined, we call a $\fsr$ oracle on $n_i,a_i$. If it returns a
nontrivial divisor $d\mid n_i$, we replace $n_i$ with $d$ and
$n_i/d$. Otherwise, it returns a square root of~$a_i$ modulo~$n_i$,
which we store as~$u_i$.
\end{itemize}
This stage terminates after $O(\lh n)$ steps. When it does, we can
write $m=\prod_{i\in I}n_i^{e_i}$ for some $e_i>0$,
$I\sset\{0,\dots,k-1\}$, where $n_i$, $i\in I$, are pairwise coprime, none of
them is a perfect square, and we have $a=n_i^{j_i}a_i$ for some~$j_i$
and $(a_i,n_i)=1$. For each~$i$, we try to compute a square root~$z_i$
of~$a$ modulo~$n_i^{e_i}$ as follows:
\begin{itemize}
\item If $j_i\ge e_i$, we put $z_i=0$.
\item If $j_i<e_i$, and $j_i$ is odd or $\tljk{a_i}{n_i}=-1$,
we return~$n_i$ as a nonsquare witness for~$a$.
\item If $j_i<e_i$ is even and $\tljk{a_i}{n_i}=1$, then $u_i$ is
defined, and $u_i^2\equiv a_i\pod{n_i}$. We put $z_i=n_i^{j_i/2}v_i$,
where $v_i^2\equiv a_i\pod{n_i^{e_i}}$ is computed using
Hensel's lifting, which is an iteration of the following procedure: if
we have $u$ such that $u^2\equiv a_i\pod{n_i^c}$, we compute
$w\equiv(2u)^{-1}\pod{n_i^c}$, and we put $u'=(u^2+a_i)w$. Then $u'^2\equiv
a_i\pod{n_i^{2c}}$.
\end{itemize}
We also try to find a square root~$z$ of~$a$ modulo $2^e$. We
write $a=2^jb$ with $b$ odd, and then:
\begin{itemize}
\item If $j\ge e$, we put $z=0$.
\item If $j<e$, we return~$2$ as a nonsquare witness for~$a$ whenever
one of the following cases happens: $j$ is odd, or
$e-j\ge2$ and $b\equiv3\pod 4$, or $e-j\ge3$ and $b\equiv5\pod 8$.
\item Otherwise, $j<e$ is even, and $1^2\equiv b\pod{2^{\min\{e-j,3\}}}$. We
put $z=2^{j/2}v$, where $v^2\equiv b\pod{2^{e-j}}$; if $e-j>3$, we
compute~$v$ using the following variant of Hensel's lifting. If
we have $u$ such that $u^2\equiv b\pod{2^c}$, we compute
$w\equiv u^{-1}\pod{2^{c-2}}$, and we put $u'=((u^2+b)/2)w$. Then $u'^2\equiv
b\pod{2^{2c-2}}$.
\end{itemize}
Finally, using the Chinese remainder theorem, we compute $x$ such that
$x\equiv z\pod{2^e}$ and $x\equiv z_i\pod{n_i^{e_i}}$ for every~$i$,
then $x^2\equiv a\pod n$.
\end{Pf}

\section{$\fsr$ is in $\ppa$}\label{sec:fsrinppa}

The purpose of this section is to prove \th\ref{thm:fsr}. As already
mentioned in the introduction, the original idea of the proof comes
from previous work of the author on the provability of the quadratic
reciprocity theorem in variants of bounded arithmetic, and in fact,
$\fsr\in\ppa$ is a simple corollary of these results. This connection
is described in detail in Section~\ref{sec:bounded-arithmetic}. In
order to make this paper more self-contained, we give a direct
combinatorial proof of \th\ref{thm:fsr} in
Section~\ref{sec:explicit-algorithm}. Readers uncomfortable with bounded
arithmetic may safely skip straight there.

\pagebreak[2]
\subsection{Bounded arithmetic}\label{sec:bounded-arithmetic}

We assume familiarity with basic facts about subsystems of
bounded arithmetic, in particular Buss's theory $\sss$. We refer the
reader to \cite{buss:ba,book} for more background.

Je\v r\'abek~\cite{ej:flt} introduced a theory $\sss+\coun_2(\pv)$,
axiomatized over~$\sss$ by the
following principle: for every number~$a$ and circuit~$C$, $C$~does not
define an involution on $\{0,\dots,2a\}$ without fixpoints. Notice
that the axiom is~$\sig1$, and the corresponding search problem is a
minor variant of~$\lone$.

\begin{Lem}\th\label{lem:wit}
If $\sss+\coun_2(\pv)\vdash\forall x\,\exists y\,\fii(x,y)$, where
$\fii\in\sig1$, then the search problem to find a~$y$ satisfying
$\fii(x,y)$ given~$x$ is in~$\ppa$.
\end{Lem}
\begin{Pf}
By the assumption, $\sss$ proves
$$\exists a,C\,\forall u\le2a\,(C(C(u))=u\ne C(u)\le2a)\lor
  \exists y\,\fii(x,y),$$
hence $\sss(h)$ proves its Herbrandization
$$\exists a,C\,(h(a,C)\le2a\to C(C(h(a,C)))=h(a,C)\ne C(h(a,C))\le2a)\lor
  \exists y\,\fii(x,y).$$
This is an $\exists\sig1(h)$ formula, hence using Parikh's theorem and
Buss's witnessing theorem, there exists a polynomial-time oracle
function~$f^h$ such that
\[\tag{$*$}\exists a,C\,(h(a,C)\le2a\to C(C(h(a,C)))=h(a,C)\ne C(h(a,C))\le2a)
 \lor\fii(x,f^h(x))\]
holds in $\stm$ for any choice of~$h$. Let us run~$f$ on an input~$x$
with an oracle solving the $\ppa$-problem corresponding to $\coun_2$
in place of~$h$, and let $y$ be its output. We may assume that $f$
never asks the same question more than once, hence the oracle answers
in any particular run can be extended to a function~$h$ which satisfies
$$h(a,C)\le2a\land
  (C(C(h(a,C)))\ne h(a,C)\lor h(a,C)=C(h(a,C))\lor C(h(a,C))>2a).$$
Then $(*)$ implies $\fii(x,y)$. Thus, the search problem associated
to $\fii$ is in $\fp^\ppa=\ppa$ using \th\ref{thm:bj}.
\end{Pf}
Let $J(a,n)$ denote a $\pv$-function formalizing the algorithm in
Figure~\ref{fig:jac}. As shown in~\cite{ej:flt}, $\sss+\coun_2(\pv)$ proves that
$J(a,n)$ agrees with the definition of the Jacobi
symbol in terms of factorization of~$n$ and quadratic residues. In
particular, the theory proves that for prime~$n$, $J(a,n)=1$ implies
that $a$ is a quadratic residue, which can be expressed as
the following $\sig1$ formula:
\begin{Thm}[Je\v r\'abek \cite{ej:flt}]
$\sss+\coun_2(\pv)$ proves
$$J(a,n)=1\to\exists x\,(x^2\equiv a\pod n)\lor\exists u,v<n\,(uv=n).\qedhere$$
\end{Thm}
\th\ref{thm:fsr} readily follows.

\pagebreak[2]
\subsection{Explicit algorithm}\label{sec:explicit-algorithm}

Before turning to $\fsr$ proper, we will describe $\ppa$ algorithms
for some of its special cases which we will need as ingredients in the
main construction.

We introduce some notation for conciseness. If $n$ is a fixed odd
integer~$n>1$, we consider
$$N=\{x:\abs x<n/2,(n,x)=1\}$$
as a set of unique representatives of $(\Z/n\Z)^*$. We also write
$N^+=\{x\in N: x>0\}$, $N^-=\{x\in N: x<0\}$, $N_0=N\cup\{0\}$, and
similarly for $N^+_0$,~$N^-_0$. We assume operations on residues are
computed modulo~$n$ with a result in~$N$, so that, e.g., $ab^{-1}\in
N^+$ means that $a\equiv bx\pod n$ for some~$x\in N^+$.
\begin{Lem}\th\label{lem:gauss}
There is a poly-time function $f(n,a,x)$ such that for any odd~$n>1$
and an integer~$a$ coprime to~$n$, the function $f_{n,a}(x)=f(n,a,x)$
defines an involution on
$$\{x\in N^-:ax\in N^-\}\cup N^+_0$$
whose fixpoints are of the form~$x^{-1}$, where
\begin{enumerate}
\item\label{item:fact} $x\in N^+\bez\{1\}$ and $x^2=1$, or
\item\label{item:root} $x\in N^-$ and $x^2=a$.
\end{enumerate}
\end{Lem}
\begin{Pf}
We define $f'_{n,a}$ on $\{x\in N^-:ax\in N^-\}\cup N^+$ by
$$f'_{n,a}(x)=\begin{cases}
x^{-1}&x,x^{-1}\in N^+,\\
a^{-1}x^{-1}&ax,x^{-1}\in N^-,\\
-x&(x,ax\in N^+\land x^{-1}\in N^-)\lor(x,ax\in N^-\land x^{-1}\in N^+).
\end{cases}
$$
It is easy to see that the three conditions define a partition of
$\{x\in N^-:ax\in N^-\}\cup N^+$, and $f'_{n,a}$ is an involution on each
part. The fixpoints of~$f'_{n,a}$ in the first two parts have the forms
\eqref{item:fact} (without the restriction $x\ne1$) and~\eqref{item:root}, respectively, and there are no fixpoints in the
third part. Finally, we put
$$f_{n,a}(x)=\begin{cases}
1&x=0,\\
0&x=1,\\
f'_{n,a}(x)&x\ne0,1.
\end{cases}\qedhere$$
\end{Pf}
\begin{Def}
For any constant~$a$, let $\fsr_a$ denote the following special case
of $\fsr$: given an odd positive~$n$ such that $\tljk an=1$, find
either a nontrivial divisor of~$n$, or a square root of~$a$ modulo~$n$.
\end{Def}
\pagebreak[2]
\begin{Lem}\th\label{lem:fsr_a}
$\fsr_{-1}$ and $\fsr_2$ are in $\ppa$.
\end{Lem}
\begin{Pf}
Given $n\equiv\pm1\pod8$, observe that
$$\{x\in N^-:2x\in N^-\}=N\cap[-(n-2\pm1)/4,-1].$$
We define an involution~$r$ on
$[-(n-2\pm1)/4,(n-1)/2]$ by
$$r(x)=\begin{cases}
x&x\ne0,(x,n)\ne1,\\
f_{n,2}(x)&\text{otherwise.}
\end{cases}$$
The domain of~$r$ is an interval of size $(3n\pm1)/4$, which is odd,
hence we can use $\lone$ to find a fixpoint $x$ of~$r$. Using
\th\ref{lem:gauss}, we see that either $x^{-1}$ is a square root
of~$2$, or it is a square root of~$1$ distinct from~$\pm1$, or
$(x,n)\ne1$. In the last two cases, we can factorize~$n$.

For $\fsr_{-1}$, we define similarly an involution on $[0,(n-1)/2]$ using
$f_{n,-1}$.
\end{Pf}
Using a similar construction, it is possible to show $\fsr_a\in\ppa$
for every constant~$a$. We skip the details, as we will not directly
need this fact, and \th\ref{thm:fsr} is more general.
However, notice that $\fsr_{-1}\in\ppa$ restates Buresh-Oppenheim's original
result, and any constant~$a$ yields a similar special case of factoring:
\begin{Exm}\th\label{exm:fsra}
The following search problems are in~$\ppa$.

Given $n\equiv\pm1\pod8$ such that $2$ is a quadratic nonresidue
modulo~$n$ (i.e., $n$ has a divisor $p\equiv\pm3\pod8$), find a
nontrivial divisor of~$n$.

Given $n\equiv1\pod3$ such that $-3$ is a quadratic nonresidue
modulo~$n$ (i.e., $n$ has a divisor $p\equiv2\pod3$), find a
nontrivial divisor of~$n$.
\end{Exm}
\begin{Lem}\th\label{lem:fsrm}
$\fsrm$ (and thus $\wfsr$) is in $\ppa$.
\end{Lem}
\begin{Pf}
Let $n>1$ be odd, and $a$,~$b$ coprime to~$n$. Define
$$g(x)=\begin{cases}
\p{0,-x}&x\in N_0^+,\\
\p{1,-x}&x,ax,b^{-1}x\in N^-,\\
\p{2,-x}&x,ax\in N^-,b^{-1}x\in N^+.
\end{cases}$$
Then $g$ is a poly-time bijection from $\{x\in N^-:ax\in N^-\}\cup
N^+_0$ onto
\begin{align*}
A=(\{0\}\times N^-_0)&\cup(\{1\}\times\{x:x,ax,b^{-1}x\in N^+\})\\
  &\cup(\{2\}\times\{x\in N^+:ax\in N^+,b^{-1}x\in N^-\})
\end{align*}
with a poly-time inverse. Similarly,
$$h(x)=\begin{cases}
\p{0,x}&x\in N^+,\\
\p{1,x}&x=0\text{ or }x,b^{-1}x,ax\in N^-,\\
\p{2,x}&x,b^{-1}x\in N^-,ax\in N^+
\end{cases}$$
is a bijection from $\{x\in N^-:b^{-1}x\in N^-\}\cup N^+_0$ onto
\begin{align*}
B=(\{0\}\times N^+)&
  \cup(\{1\}\times(\{0\}\cup\{x:x,ax,b^{-1}x\in N^-\}))\\
  &\cup(\{2\}\times\{x\in N^-:ax\in N^+,b^{-1}x\in N^-\}),
\end{align*}
$x\mapsto\p{2,bx}$ is a bijection from $\{x\in N^-:abx\in
N^-\}\cup N^+_0$ onto
$$C=\{2\}\times(\{0\}\cup\{x:ax\in N^-\lor b^{-1}x\in N^+\}),$$
and $\p{1,x}\mapsto\p{1,-x}$ is a fixpoint-free involution on
$$D=\{1\}\times\{x\in N:x,ax,b^{-1}x\text{ do not have the same sign}\}.$$
We can thus define a poly-time involution~$r$ on
$\{0,1,2\}\times[-(n-1)/2,(n-1)/2]$ by
$$r(e,x)=\begin{cases}
g(f_{n,a}(g^{-1}(e,x)))&\p{e,x}\in A,\\
h(f_{n,b^{-1}}(h^{-1}(e,x)))&\p{e,x}\in B,\\
\p{2,bf_{n,ab}(b^{-1}x)}&\p{e,x}\in C,\\
\p{1,-x}&\p{e,x}\in D,\\
\p{e,x}&x\ne0,(x,n)>1.
\end{cases}$$
Since $3n$ is odd, we can use $\lone$ to find a fixpoint $\p{e,x}$
of~$r$. We cannot have $\p{e,x}\in D$. If $x\ne0$, $(x,n)>1$, we can
factor~$n$. If $\p{e,x}\in A$, then $y:=g^{-1}(e,x)$ is a fixpoint of
$f_{n,a}$. Thus, either $y^2=1$, $y\ne\pm1$, in which case we can
factor~$n$, or $y^{-1}$ is a square root of~$a$. Similarly, if
$\p{e,x}\in B\cup C$, we can factor~$n$, or compute a square root
of $b$ or~$ab$.
\end{Pf}

\th\ref{lem:fsrm} is enough to prove our main result,
\th\ref{thm:fac}. However, we will proceed with the proof of
\th\ref{thm:fsr}, as we are interested in the possibility of
unconditional derandomization of the reduction of factoring to $\ppa$,
and placing $\fsr$ in $\ppa$ can be seen as a partial step towards
that goal. Moreover, randomized versions of
\th\ref{prop:nonjac,thm:root} would not be interesting.

\begin{Lem}\th\label{lem:evenodd}
The following problems are in~$\ppa$.
\begin{enumerate}
\item\label{item:odd}
$\fsro$: given an odd~$n>0$, a sequence $\p{a_i:i<k}$ of integers
coprime to~$n$ such that $k$ is odd, and a square root $x$
of\/~$\prod_{i<k}a_i$ modulo~$n$, find a nontrivial divisor of~$n$, or a
square root of some~$a_i$ modulo~$n$.
\item\label{item:even}
$\fsre$: given an odd~$n>0$, and a sequence $\p{a_i:i<k}$ of integers
coprime to~$n$ such that $k$ is even, find a nontrivial divisor of~$n$, or a
square root of\/~$\prod_{i<k}a_i$ or of some~$a_i$ modulo~$n$.
\end{enumerate}
\end{Lem}
\begin{Pf}
\eqref{item:odd}:
Put $I=\{0,\dots,k-1\}$ and~$y=1$, and repeat the following steps. If
$I=\{i\}$, return~$xy^{-1}$ as a square root of~$a_i$. If $\abs I>1$, pick
$i,j\in I$, $i\ne j$, and call $\fsrm$ on~$n,a_i,a_j$. If it gives us a
nontrivial divisor of~$n$, or a square root of $a_i$ or~$a_j$, we
return it. Otherwise, it provides a square root $z$ of~$a_ia_j$. We
multiply $y$ by~$z$, remove $i,j$ from~$I$, and repeat the loop.

\eqref{item:even}: Put $x=a_k=\prod_{i<k}a_i$, and call $\fsro$.
\end{Pf}

\begin{Def}
$\qrec$ is the following problem: given odd coprime $n,m>0$ such that
$n\equiv1\pod4$, and a square root~$a$ of~$n$ modulo~$m$, find a
nontrivial divisor of $n$ or~$m$, or a square root of~$m$ modulo~$n$.

Notice that $\qrec$ is a special case of $\fsr$: the input data ensure
$\tljk nm=1$, hence $\tljk mn=1$ by quadratic reciprocity.
\end{Def}
\begin{Lem}\th\label{lem:wqrec}
$\qrec\in\ppa$.
\end{Lem}
\begin{Pf}
We may assume $n,m>1$ and $a\in M^-$, so that $b=a^{-1}\in
M$ is a fixpoint of~$f_{m,n}$. Put $n_2=(n+1)/2$, $m_2=(m+1)/2$. The function
$$g(x)=\begin{cases}
\p{x,m_2}&x\in N^+_0,\\
\p{-x,\fl{-mx/n}}&x,mx\in N^-
\end{cases}$$
is a poly-time bijection with poly-time inverse from $\{x\in N^-:mx\in
N^-\}\cup N^+_0$ onto
$$A=\bigl(N^+_0\times\{m_2\}\bigr)\cup
  \bigl\{\p{x,y}\in N^+_0\times[0,m_2):mx-ny\in N^+\bigr\},$$
where $mx-ny$ is \emph{not} evaluated modulo~$n$, but literally. Likewise,
$$h(y)=\begin{cases}
\p{n_2,y}&y\in M^+_0,\\
\p{\fl{-ny/m},-y}&y,ny\in M^-
\end{cases}$$
is a bijection from $\{y\in M^-:ny\in M^-\}\cup M^+_0$ onto
$$B=\bigl(\{n_2\}\times M^+_0\bigr)\cup
  \bigl\{\p{x,y}\in[0,n_2)\times M^+_0:mx-ny\in M^-\bigr\}.$$
The function $k(x,y)=\p{n_2-1-x,m_2-1-y}$ is a poly-time involution
with no fixpoints on
$$C=\bigl\{\p{x,y}\in[0,n_2)\times[0,m_2):
         mx-ny\ge n_2\text{ or }mx-ny\le-m_2\bigr\}.$$
We define a poly-time involution~$r$ on
$([0,n_2]\times[0,m_2])\bez\{\p{n_2,m_2}\}$ by
$$r(x,y)=\begin{cases}
g(f_{n,m}(g^{-1}(x,y)))&\p{x,y}\in A,\\
h(f_{m,n}(h^{-1}(x,y)))&\p{x,y}\in B\bez\{h(b)\},\\
\p{0,0}&\p{x,y}=h(b),\\
h(b)&x=y=0,\\
k(x,y)&\p{x,y}\in C,\\
\p{x,y}&\text{otherwise.}
\end{cases}$$
Notice that if $x\in[0,n_2)$ and $y\in[0,m_2)$ are such that
$mx-ny=0$, then $x=y=0$, as $(n,m)=1$. It follows that the last clause
in the definition of~$r$ applies to elements of the set
\begin{multline*}
D=\bigl(([1,n_2)\bez N^+)\times\{m_2\}\bigr)
  \cup\bigl(\{n_2\}\times([1,m_2)\bez M^+)\bigr)\\
  \cup\bigl\{\p{x,y}\in[0,n_2)\times[0,m_2):
      mx-ny\in([1,n_2)\bez N^+)\cup((-m_2,-1]\bez M^-)\bigr\}.
\end{multline*}
The domain of~$r$ has odd size $(n_2+1)(m_2+1)-1$, hence using
$\lone$, we can find a fixpoint $\p{x,y}$ of~$r$. If $\p{x,y}\in A$,
it gives us a square root of~$m$ modulo~$n$, or a square root of~$1$
distinct from~$\pm1$, in which case we can factorize~$n$. If
$\p{x,y}\in B$, we get a square root of~$n$ modulo~$m$ distinct
from~$\pm a$, or a square root of~$1$ distinct from~$\pm1$, and both
cases give a factor of~$m$. If $\p{x,y}\in D$, $(n,x)$ or~$(m,y)$ is a
nontrivial divisor of $n$ or~$m$, respectively.
\end{Pf}

We are ready now to prove \th\ref{thm:fsr}. Assume we are given an
odd~$n>0$, and an integer~$a$ such that $\tljk an=1$. We first compute
the sequences $\p{a_i:i\le t}$, $\p{n_i:i\le t}$ of values of $a$
and~$n$ during the execution of the algorithm in Figure~\ref{fig:jac}.
That is, we put $\p{a_0,n_0}=\p{a,n}$, and then we define
$\p{a_i,n_i}$ by induction on~$i$ as follows. If $\abs{a_i}>n_i/2$, we let
$n_{i+1}=n_i$, and $a_{i+1}\equiv a_i\pod{n_i}$ such that
$\abs{a_{i+1}}<n_i/2$. If $0<\abs{a_i}<n_i/2$, we define
$$\p{a_{i+1},n_{i+1}}=\begin{cases}
\p{-a_i,n_i}&a_i<0,\\
\p{a_i/2,n_i}&a_i>0\text{ is even,}\\
\p{n_i,a_i}&a_i>0\text{ is odd.}
\end{cases}$$
We stop when we reach $a_t=0$. Since $\tljk an=1$, we have
$(a_i,n_i)=1$ for each~$i$, in particular $n_t=1$. Notice that
$t=O(\lh n)$. Write $R=\{i<t:a_i\text{ is odd},0<a_i<n_i/2\}$.

In the main part of the algorithm, we maintain a double sequence
$\p{n_{i,j}:i\le t,j<s_i}$ of integers~$n_{i,j}>1$ such that
$n_i=\prod_{j<s_i}n_{i,j}$, and $n_{i,j}\le n_{i,j'}$ for~$j<j'$.
Moreover, we maintain sequences $\p{u_{i,j}:i\le k,j<s_i}$,
$\p{v_{i,j,k},w_{i,j,k}:i\in R,j<s_i,k<s_{i+1}}$, where some of the
$u_{i,j}$,~$v_{i,j,k}$, and~$w_{i,j,k}$ may be undefined. Where they
are defined, we have $u_{i,j}^2\equiv a_i\pod{n_{i,j}}$,
$v_{i,j,k}^2\equiv n_{i+1,k}\pod{n_{i,j}}$, and $w_{i,j,k}^2\equiv
n_{i,j}\pod{n_{i+1,k}}$, respectively.

We initialize the sequences with~$s_i=1$, $n_{i,0}=n_i$ for~$n_i>1$,
$s_i=0$ for~$n_i=1$, and all $u_{i,j}$, $v_{i,j,k}$, and~$w_{i,j,k}$
undefined. We repeat in arbitrary order the following updating steps
until none of them is applicable any more.
\begin{itemize}
\item Assume $n_i=n_{i+1}$, $n_{i,j}\ne n_{i+1,k}$, and
$d=(n_{i,j},n_{i+1,k})>1$. If $d\ne n_{i,j}$, we increase~$s_i$, replace $n_{i,j}$
with~$d$ and~$n_{i,j}/d$, and undefine all associated
$u_{i,j}$,~$v_{i-1,l,j}$, and~$w_{i-1,l,j}$. If $d\ne n_{i+1,k}$, we
deal with it similarly. Notice that we cannot have
$n_{i,j}=d=n_{i+1,k}$.

Moreover, if this step is not applicable, then
$n_i=n_{i+1}$ implies that $s_i=s_{i+1}$ and $\p{n_{i,j}:j<s_i}$ and
$\p{n_{i+1,k}:k<s_{i+1}}$ are permutations of each other, hence in view of
their monotonicity, we have $n_{i,j}=n_{i+1,j}$ for each~$j$.
\item For $i<t$ such that $\p{n_{i,j}:j<s_i}=\p{n_{i+1,k}:k<s_{i+1}}$
(which implies $n_i=n_{i+1}$):
\begin{itemize}
\item If $a_i\equiv a_{i+1}\pod{n_i}$, and exactly one of $u_{i,j}$,
$u_{i+1,j}$ is defined, we define the other to the same value.
\item If $a_i=\alpha a_{i+1}$, $\alpha\in\{-1,2\}$,
$\tljk\alpha{n_{i,j}}=-1$, and neither~$u_{i,j}$ nor~$u_{i+1,j}$ is
defined, we call $\fsrm(n_{i,j},a_i,a_{i+1})$. If it returns a nontrivial
divisor of~$n_{i,j}$, we expand the $n_{i,j}$ sequence as in the first
step. Otherwise, it gives a square root of $a_i$ or~$a_{i+1}$
modulo~$n_{i,j}$, which we store as $u_{i,j}$ or~$u_{i+1,j}$, respectively.
\item If $a_i=\alpha a_{i+1}$, $\alpha\in\{-1,2\}$,
$\tljk\alpha{n_{i,j}}=1$, and exactly one of $u_{i,j}$ or~$u_{i+1,j}$ is
defined, we call $\fsr_\alpha(n_{i,j})$. If it returns a nontrivial
divisor of~$n_{i,j}$, we expand the $n_{i,j}$ sequence. Otherwise, it
gives $\beta^2\equiv\alpha\pod{n_{i,j}}$, and we define
$u_{i,j}:=\beta u_{i+1,j}$ or $u_{i+1,j}:=\beta^{-1}u_{i,j}$, respectively.
\end{itemize}
\item For $i\in R$:
\begin{itemize}
\item If $u_{i,j}$ is defined and $\abs I$ is odd, where
$I=\{k<s_{i+1}:v_{i,j,k}\text{ is undefined}\}$, we put
$x=u_{i,j}\prod_{k\notin I}v_{i,j,k}^{-1}$, and call
$\fsro$ on~$n_{i,j},\p{n_{i+1,k}:k\in I},x$. If it returns a factor
of~$n_{i,j}$, we expand the $n_{i,j}$ sequence. Otherwise, it returns
a square root of some~$n_{i+1,k}$, $k\in I$, modulo~$n_{i,j}$, which we
store as~$v_{i,j,k}$.
\item If $u_{i,j}$ is undefined and $\abs I$ is even, where
$I$ is as above, we call
$\fsre$ on~$n_{i,j},\p{n_{i+1,k}:k\in I}$. If it returns a factor
of~$n_{i,j}$, we expand the $n_{i,j}$ sequence. If it returns
a square root of some~$n_{i+1,k}$, $k\in I$, modulo~$n_{i,j}$, we
store it as~$v_{i,j,k}$. Otherwise, it returns a square root $x$
of~$\prod_{k\in I}n_{i+1,k}$, and then we define $u_{i,j}=x\prod_{k\notin
I}v_{i,j,k}$.
\item If $u_{i+1,k}$ is defined and $\abs I$ is odd, or $u_{i+1,k}$
is undefined and $\abs I$ is even, where $I=\{j<s_i:w_{i,j,k}\text{ is
undefined}\}$, we proceed in a similar way to expand the $n_{i+1,k}$
sequence or to define some~$w_{i,j,k}$ or~$u_{i+1,k}$.
\item If $n_{i,j}\equiv-1\pod4$, $\tljk{n_{i+1,k}}{n_{i,j}}=1$, and
$v_{i,j,k}$ is undefined, we call $\wfsr$ on $n_{i,j},n_{i+1,k},-1$. If it
returns a factor of~$n_{i,j}$, we expand the $n_{i,j}$ sequence,
otherwise it returns a square root of~$n_{i+1,k}$ modulo~$n_{i,j}$,
which we store as~$v_{i,j,k}$.
\item If $n_{i+1,k}\equiv-1\pod4$, $\tljk{n_{i,j}}{n_{i+1,k}}=1$, and
$w_{i,j,k}$ is undefined, we proceed similarly.
\item If $n_{i,j}\equiv1\pod4$, $w_{i,j,k}$ is defined, and
$v_{i,j,k}$ is undefined, we call
$\qrec$ on $n_{i,j},n_{i+1,k},w_{i,j,k}$.  If it
returns a factor of $n_{i,j}$ or~$n_{i+1,k}$, we expand the $n_{i,j}$
or $n_{i+1,k}$ sequence (respectively),
otherwise it returns a square root of~$n_{i+1,k}$ modulo~$n_{i,j}$,
which we store as~$v_{i,j,k}$.
\item If $n_{i+1,k}\equiv1\pod4$, $v_{i,j,k}$ is defined, and
$w_{i,j,k}$ is undefined, we proceed similarly.
\end{itemize}
\end{itemize}
In each step, either $\sum_{i\le t}s_i\le O(\lh n^2)$ strictly increases, or it
stays the same, and we define some previously undefined
$u_{i,j}$,~$v_{i,j,k}$, or~$w_{i,j,k}$. It follows that the update
procedure stops after $\lh n^{O(1)}$ steps.

Let us write $\tmljk{a_i}{n_{i,j}}=1$ if $u_{i,j}$ is defined, and
$\tmljk{a_i}{n_{i,j}}=-1$ otherwise. We define
$\tmljk{n_{i+1,k}}{n_{i,j}}$ and~$\tmljk{n_{i,j}}{n_{i+1,k}}$
similarly using $v_{i,j,k}$ and~$w_{i,j,k}$, respectively. Notice that
$\tmljk{a_i}{n_{i,j}}=1$ implies $\tljk{a_i}{n_{i,j}}=1$, and
likewise for $\tmljk{n_{i+1,k}}{n_{i,j}}$, $\tmljk{n_{i,j}}{n_{i+1,k}}$.
\begin{Lem}\th\label{lem:res}
When the update procedure stops, the following properties hold.
\begin{enumerate}
\item\label{item:eqeq}
If $n_i=n_{i+1}$, then $s_i=s_{i+1}$ and $n_{i,j}=n_{i+1,j}$.
\item\label{item:equiv}
If $n_i=n_{i+1}$ and $a_i\equiv a_{i+1}\pod{n_i}$, then
$\tmljk{a_i}{n_{i,j}}=\tmljk{a_{i+1}}{n_{i+1,j}}$.
\item\label{item:2-1}
If $n_i=n_{i+1}$ and $a_i=\alpha a_{i+1}$, $\alpha\in\{-1,2\}$, then
$$\mljk{a_{i+1}}{n_{i+1,j}}=\ljk\alpha{n_{i,j}}\mljk{a_i}{n_{i,j}}.$$
\item\label{item:prod}
If $i\in R$, then
$$\mljk{a_i}{n_{i,j}}=\prod_{k<s_{i+1}}\mljk{n_{i+1,k}}{n_{i,j}},
\qquad\mljk{a_{i+1}}{n_{i+1,k}}=\prod_{j<s_i}\mljk{n_{i,j}}{n_{i+1,k}}.$$
\item\label{item:qrec-1}
If $i\in R$ and $n_{i,j}\equiv n_{i+1,k}\equiv-1\pod4$, then
$$\mljk{n_{i+1,k}}{n_{i,j}}\mljk{n_{i,j}}{n_{i+1,k}}=-1.$$
\item\label{item:qrec1}
If $i\in R$ and $n_{i,j}\equiv1\pod4$ or $n_{i+1,k}\equiv1\pod4$, then
$$\mljk{n_{i+1,k}}{n_{i,j}}\mljk{n_{i,j}}{n_{i+1,k}}=1.$$
\end{enumerate}
\end{Lem}
\begin{Pf}
\eqref{item:eqeq},~\eqref{item:equiv}, and~\eqref{item:prod} are clear.

\eqref{item:2-1}: The statement is clear if $\tljk\alpha{n_{i,j}}=1$.
If $\tljk\alpha{n_{i,j}}=-1$, the inapplicability of update steps
implies that $\tmljk{a_i}{n_{i,j}}=1$
or~$\tmljk{a_{i+1}}{n_{i+1,j}}=1$. We cannot have both, since this
would imply $\tljk{a_i}{n_{i,j}}=\tljk{a_{i+1}}{n_{i,j}}=1$, hence $\tljk\alpha{n_{i,j}}=1$.

\eqref{item:qrec-1}: By quadratic reciprocity, exactly one
of~$\tljk{n_{i+1,k}}{n_{i,j}}=1$, $\tljk{n_{i,j}}{n_{i+1,k}}=1$ holds.
The inapplicability of update steps then implies that
$\tmljk{n_{i+1,k}}{n_{i,j}}=1$ or $\tmljk{n_{i,j}}{n_{i+1,k}}=1$. We
cannot have both, as this would mean
$\tljk{n_{i+1,k}}{n_{i,j}}=\tljk{n_{i,j}}{n_{i+1,k}}=1$.

\eqref{item:qrec1}: The statement is clear if $n_{i,j}\equiv
n_{i+1,k}\equiv1\pod4$. Assume $n_{i,j}\equiv1\pod4$ and
$n_{i+1,k}\equiv-1\pod4$, the other case is symmetric. By the
inapplicability of update steps, $\tmljk{n_{i,j}}{n_{i+1,k}}=1$
implies $\tmljk{n_{i+1,k}}{n_{i,j}}=1$. On the other hand, if
$\tmljk{n_{i+1,k}}{n_{i,j}}=1$, then $\tljk{n_{i+1,k}}{n_{i,j}}=1$,
hence $\tljk{n_{i,j}}{n_{i+1,k}}=1$ by quadratic reciprocity, thus
$\tmljk{n_{i,j}}{n_{i+1,k}}=1$ by the inapplicability of update steps.
\end{Pf}
Using \th\ref{lem:res}, we can show
$$\prod_{j<s_i}\mljk{a_i}{n_{i,j}}=\ljk{a_i}{n_i}$$
by reverse induction on~$i$. The induction step for $i\in R$ goes as
follows:
\begin{align*}
\prod_{j<s_i}\mljk{a_i}{n_{i,j}}
&=\prod_{\substack{j<s_i\\k<s_{i+1}}}\mljk{n_{i+1,k}}{n_{i,j}}\\
&=\prod_{\substack{j<s_i\\k<s_{i+1}}}
    \mljk{n_{i,j}}{n_{i+1,k}}(-1)^{(n_{i,j}-1)(n_{i+1,k}-1)/4}\\
&=(-1)^{(a_{i+1}-1)(n_{i+1}-1)/4}\prod_{k<s_{i+1}}\mljk{a_{i+1}}{n_{i+1,k}}\\
&=(-1)^{(a_{i+1}-1)(n_{i+1}-1)/4}\ljk{a_{i+1}}{n_{i+1}}\\
&=\ljk{n_{i+1}}{a_{i+1}}=\ljk{a_i}{n_i}.
\end{align*}

In particular, either $s_0>1$, in which case $n_{0,0}$ is a nontrivial divisor
of~$n$, or $s_0=1$ and~$\tmljk{a_0}{n_{0,0}}=1$, where $a_0=a$
and~$n_{0,0}=n$, in which case $u_{0,0}^2\equiv a\pod n$. This
completes the proof of \th\ref{thm:fsr}.

\section{Conclusion}\label{sec:conclusion}

We have shown that integer factoring has randomized reductions to
the classes $\ppa$ and~$\ppp$ (more precisely,
$\pwpp$). We also provided evidence that there in fact exist
deterministic reductions, namely this is true under the widely believed
assumption of the generalized Riemann hypothesis for quadratic
Dirichlet characters.
\begin{Prob}\th\label{prob:derand}
Is $\fac$ in $\ppa$, $\ppp$, or~$\fp^\ppp$?
\end{Prob}
Some of our other results can be seen as partial indication that such an
unconditional deterministic reduction might be possible at least in the
case of~$\ppa$. In particular, the fact that $\fsr\in\ppa$ bypasses
the randomized reduction of~$\wfsr$ to~$\fsr$, and we have shown that
$\ppa$ contains the search problems to find square roots modulo
arbitrary integers (which is probabilistically Turing-equivalent to
factoring) and to find quadratic nonresidues (which is easily solvable
in randomized polynomial time). Nevertheless, it remains open whether
\th\ref{prob:derand} can be resolved unconditionally.

Another interesting question is whether the methods used for the
reduction of factoring to~$\ppa$ can be pushed down to the class
$\ppad\sset\ppa$. Note that many natural problems are known to be
complete for~$\ppad$, such as computing Nash equilibria \cite{nashppad}.
\begin{Prob}\th\label{prob:ppad}
Does $\fac$ have some form of reduction to $\ppad$?
\end{Prob}

\subsection*{Acknowledgements}
I would like to thank Josh Buresh-Oppenheim for a clarification of his
work, and Rahul Savani for a useful suggestion.

\bibliographystyle{mybib}
\bibliography{mybib}

\providecommand\gobble[1]{} {\catcode`\/=13
  \gdef/{\string/\futurelet\nexttoken\finishslash}
  \gdef\finishslash{\ifx\nexttoken/\else\penalty\relpenalty\fi}}
  \providecommand\url{\begingroup\catcode`\~=12 \catcode`\/=13 \finishurl}
  \def\finishurl#1{\texttt{#1}\endgroup}
  \providecommand\dotminus{\mathbin{\scriptstyle\dot{\smash{\textstyle-}}}}
\providecommand{\bysame}{\leavevmode\hbox to5em{\hrulefill}\thinspace}
\providecommand\bibliographyhook{}
\begin{thebibliography}{10}
\bibliographyhook

\bibitem{aks:prime}
Manindra Agrawal, Neeraj Kayal, and Nitin Saxena, \emph{{$\mathrm{PRIMES}$} is
  in {$\mathrm P$}}, Annals of Mathematics 160 (2004), no.~2, pp.~781--793.

\bibitem{ank}
Nesmith~C. Ankeny, \emph{The least quadratic non residue}, Annals of
  Mathematics 55 (1952), no.~1, pp.~65--72.

\bibitem{bach}
Eric Bach, \emph{Explicit bounds for primality testing and related problems},
  Mathematics of Computation 55 (1990), no.~191, pp.~355--380.

\bibitem{bceip}
Paul Beame, Stephen Cook, Jeff Edmonds, Russell Impagliazzo, and Toniann
  Pitassi, \emph{The relative complexity of\/ {$\mathsf{NP}$} search problems},
  Journal of Computer and System Sciences 57 (1998), no.~1, pp.~3--19.

\bibitem{berint}
Alessandro Berarducci and Benedetto Intrigila, \emph{Combinatorial principles
  in elementary number theory}, Annals of Pure and Applied Logic 55 (1991),
  no.~1, pp.~35--50.

\bibitem{jbo:comm}
Joshua Buresh-Oppenheim, private communication.

\bibitem{jbo}
\bysame, \emph{On the {$\mathbf{TFNP}$} complexity of factoring}, unpublished
  note, \url{http://www.cs.toronto.edu/~bureshop/factor.pdf}, 2006.

\bibitem{buss:ba}
Samuel~R. Buss, \emph{First-order proof theory of arithmetic}, in: Handbook of
  Proof Theory (S.~R. Buss, ed.), Studies in Logic and the Foundations of
  Mathematics vol. 137, Elsevier, Amsterdam, 1998, pp.~79--147.

\bibitem{buss-john}
Samuel~R. Buss and Alan~S. Johnson, \emph{Propositional proofs and reductions
  between {$\mathrm{NP}$} search problems}, Annals of Pure and Applied Logic
  163 (2012), no.~9, pp.~1163--1182.

\bibitem{nashppad}
Xi~Chen and Xiaotie Deng, \emph{Settling the complexity of two-player {N}ash
  equilibrium}, in: Proceedings of the 47th {A}nnual {IEEE} {S}ymposium on
  {F}oundations of {C}omputer {S}cience, 2006, pp.~261--271.

\bibitem{ej:wphpvar}
Emil Je{\v r}{\'a}bek, \emph{On independence of variants of the weak pigeonhole
  principle}, Journal of Logic and Computation 17 (2007{\gobble c}), no.~3,
  pp.~587--604.

\bibitem{ej:flt}
\bysame, \emph{Abelian groups and quadratic residues in weak arithmetic},
  Mathematical Logic Quarterly 56 (2010{\gobble a}), no.~3, pp.~262--278.

\bibitem{book}
Jan Kraj{\'\i}{\v c}ek, \emph{Bounded arithmetic, propositional logic, and
  complexity theory}, Encyclopedia of Mathematics and Its Applications vol.~60,
  Cambridge University Press, 1995.

\bibitem{papa:parity}
Christos~H. Papadimitriou, \emph{On the complexity of the parity argument and
  other inefficient proofs of existence}, Journal of Computer and System
  Sciences 48 (1994), no.~3, pp.~498--532.

\bibitem{rab:cryp}
Michael~O. Rabin, \emph{Digitalized signatures and public-key functions as
  intractable as factorization}, Technical Report MIT/LCS/TR-212, MIT
  Laboratory for Computer Science, 1979.

\end{thebibliography}
\end{document}